\documentclass[twocolumn]{aastex63}
\usepackage[utf8]{inputenc}
\usepackage{amsmath}
\usepackage{xcolor}
\usepackage{appendix}
\usepackage{longtable}
\usepackage{tablefootnote}
\usepackage{comment}

\newcolumntype{?}{!{\vrule width 4pt}}

\begin{document}

\title{A Tale of Two Peas-In-A-Pod: The Kepler-323 and Kepler-104 Systems}
\author[0009-0007-6386-151X]{C. Alexander Thomas}
\affiliation{Department of Physics and Astronomy, University of Notre Dame, Notre Dame, IN 46556, USA}
\author[0000-0002-3725-3058]{Lauren M. Weiss}
\affiliation{Department of Physics and Astronomy, University of Notre Dame, Notre Dame, IN 46556, USA}

\author[0000-0002-0531-1073]{Howard Isaacson}
\affiliation{Department of Astronomy,  University of California Berkeley, Berkeley CA 94720, USA}
\affiliation{Centre for Astrophysics, University of Southern Queensland, Toowoomba, QLD, Australia}

\author[0000-0002-0298-8089]{Hilke E. Schlichting}
\affiliation{Department of Earth, Planetary, and Space Sciences, The University of California, Los Angeles, 595 Charles E. Young Drive East, Los Angeles, CA 90095, USA}

\author[0000-0001-7708-2364]{Corey Beard}
\altaffiliation{NASA FINESST Fellow}
\affiliation{Department of Physics \& Astronomy, The University of California, Irvine, Irvine, CA 92697, USA}

\author[0000-0002-4480-310X]{Casey L. Brinkman}
\affiliation{Institute for Astronomy, University of Hawai'i, 2680 Woodlawn Drive, Honolulu, HI 96822 USA}

\author[0000-0003-1125-2564]{Ashley Chontos}
\altaffiliation{Henry Norris Russell Fellow}
\affiliation{Department of Astrophysical Sciences, Princeton University, 4 Ivy Lane, Princeton, NJ 08540, USA}
\affiliation{Institute for Astronomy, University of Hawai`i, 2680 Woodlawn Drive, Honolulu, HI 96822, USA}

\author[0000-0002-4297-5506]{Paul Dalba}
\affiliation{Department of Earth \& Planetary Sciences, University of California Riverside, 900 University Ave, Riverside, CA 92521, USA}

\author[0000-0002-8958-0683]{Fei Dai}
\altaffiliation{NASA Sagan Fellow}
\affiliation{Division of Geological and Planetary Sciences,
1200 E California Blvd, Pasadena, CA, 91125, USA}
\affiliation{Department of Astronomy, California Institute of Technology, Pasadena, CA 91125, USA}

\author[0000-0002-8965-3969]{Steven Giacalone}
\affiliation{Department of Astronomy,  University of California Berkeley, Berkeley CA 94720, USA}
\affiliation{Department of Astronomy, California Institute of Technology, Pasadena, CA 91125, USA}

\author[0000-0001-8342-7736]{Jack Lubin}
\affiliation{Department of Physics \& Astronomy, The University of California Irvine, Irvine, CA 92697, USA}

\author[0000-0002-4290-6826]{Judah Van Zandt}
\affiliation{Department of Physics \& Astronomy, University of California Los Angeles, Los Angeles, CA 90095, USA}

\author[0000-0002-7670-670X]{Malena Rice}
\affiliation{Department of Astronomy, Yale University, New Haven, CT 06511, USA}

\begin{abstract}

In order to understand the relationship between planet multiplicity, mass, and composition, we present newly measured masses of five planets in two planetary systems: Kepler-323 and Kepler-104. We used the HIRES instrument at the W.M. Keck Observatory to collect 79 new radial velocity measurements (RVs) for Kepler-323, which we combined with 48 literature RVs from TNG/HARPS-N. We also conducted a reanalysis of the Kepler-104 system, using 44 previously published RV measurements. Kepler-323 b and c have masses of $2.0^{+1.2}_{-1.1}$ M$_\Earth$ and 6.5$\pm1.6$ M$_\Earth$, respectively, whereas the three Kepler-104 planets are more massive (10.0$\pm2.8$ M$_\Earth$, $7.1^{+3.8}_{-3.5}$ M$_\Earth$, and $5.5^{+4.6}_{-3.5}$ M$_\Earth$ for planets b, c, and d, respectively).  The Kepler-104 planets have densities consistent with rocky cores overlaid with gaseous envelopes ($4.1^{+1.2}_{-1.1}$ g/cc, $2.9^{+1.7}_{-1.5}$ g/cc, and $1.6^{+1.5}_{-1.1}$ g/cc respectively), whereas the Kepler-323 planets are consistent with having rocky compositions ($4.5^{+2.8}_{-2.4}$ g/cc and $9.9^{+2.7}_{-2.5}$ g/cc).  The Kepler-104 system has among the lowest values for gap complexity ($\mathcal{C}$ = 0.004) and mass partitioning ($\mathcal{Q}$ = 0.03); whereas, the Kepler-323 planets have a mass partitioning similar to that of the Inner Solar System ($\mathcal{Q}$ = 0.28 and $\mathcal{Q}$ = 0.24, respectively). For both exoplanet systems, the uncertainty in the mass partitioning is affected equally by (1) individual mass errors of the planets and (2) the possible existence of undetected low-mass planets, meaning that both improved mass characterization and improved sensitivity to low-mass planets in these systems would better elucidate the mass distribution among the planets.
\end{abstract}


\section{Introduction}

Of the 190,000 stars observed as part of the \textit{Kepler} Mission, approximately 25\% host multiple transiting planets with orbital periods less than 100 days, suggesting 30-80\% of Kepler stars host multiple planets \citep{Zhu2018,Mulders2018}.  One intriguing finding is that the planets in \textit{Kepler}'s multi-planet systems tend to be similar in size and regularly spaced, like ``peas-in-a-pod'' \citep{Weiss2018}.  Measuring the masses of planets in peas-in-a-pod like systems allows us to address several overarching questions about the nature of multi-planet systems.  (1) What are the masses and bulk compositions of the individual planets? (2) Are the masses and radii of the planets typical for their orbital periods? (3) Within a given system, what is the mass diversity and spacing distribution of transiting planets? (4) How could the presence of undetected, low-mass planets affect the statistics of the intra-system size and mass uniformity?

A major advantage of characterizing transiting exoplanets is that the planet radii and orbital periods are typically well-determined from transits, while the host star properties can be characterized through spectroscopic, imaging, and astrometric follow-up.  In this paper, we investigate two systems of multiple transiting planets that were discovered as part of the NASA \textit{Kepler} Mission: Kepler-323 and Kepler-104. In terms of their size and spacing, both Kepler-323 and Kepler-104 follow the peas-in-a-pod pattern, making them ideal test-beds for the questions posed above. However, a key physical property that is not readily determined from photometry alone is the planet masses. We selected Kepler-323 and Kepler-104 as part of a multi-semester, magnitude-limited radial velocity (RV) survey to measure planet masses in multi-planet systems. In this paper, we present new measurements of the Kepler-323 system and conduct an analysis of the properties of the Kepler-323 and Kepler-104 systems. 


\subsection{Kepler-323}
Kepler-323 is a sun-like star (0.95 $\pm$ 0.03 M$_\odot$, 1.09 $\pm$0.29 R$_\odot$). The system consists of two Earth-sized planets, 1.35 $\pm$ 0.04 R$_\Earth$ and 1.53 $\pm$ 0.05 R$_\Earth$, in short orbits of 1.67 days and 3.55 days respectively \citep{CKS7}. \citet{Bonomo2023} collected 48 radial velocities (RVs) of Kepler-323 with the HARPS-N spectrograph on the 3.6m Telescopio Nazionale Galileo at the Observatorio Roque de Los Muchachos in in La Plama, Spain. In this work, we combine those RVs with 79 new RVs from the HIRES spectrograph on the W. M. Keck Observatory 10m telescope Keck I on Maunakea, Hawaii \citep{Vogt1994} to better characterize the Kepler-323.

\subsection{Kepler-104} 
Kepler-104 is another Sun-like star ($0.82\pm0.01$ M$_\odot$, 1.05 $\pm0.03$ R$_\odot$). Its planetary system consists of three sub-Neptunes, R$_b$ = 2.38 $\pm$ 0.06 R$_\Earth$, R$_c$ = 2.36 $\pm$ 0.08 R$_\Earth$, and R$_d=$ 2.64 $\pm$ 0.14 R$_\Earth$ at orbital periods of 11.43, 23.67, and 51.76 days \citep{CKS7}. Kepler-104 has a stellar companion reported in \citet{Lillo-Box2014} and \citet{Furlan2017}, at an angular separation of approximately 1.8\arcsec. There is also a companion at an angular separation of 17\arcsec\ identified based on \textit{GAIA} data. \citep{Mugrauer2019}. For an angular separation of 1.8\arcsec, a distance of 400 pc \citep{GAIADR3} corresponds to a semi-major axis of 720 AU and an orbital period of 20,000 years, and so we do not expect this companion to affect our analysis. We leverage 44 published RVs of Kepler-104 from \citet{KGPS} to provide a detailed analysis of the mass diversity of the system.

\section{New Observations}
We collected 79 spectra of Kepler-323 using HIRES between May 20th, 2019 and October 17th, 2021. To measure RVs, we used the standard California Planet Search (CPS) data reduction pipeline as described in Howard et al. (2010). This method involves using a warm iodine cell mounted in front of the slit to imprint absorption features at reference wavelengths determined from a high-resolution atlas of iodine lines \citep{Butler1996}.  Since clouds and/or moonlight can contaminate the spectra of stars with $V>10.5$ \citep{Marcy2014}, we gathered observations in a manner that enabled sky subtraction.  Specifically, observations were taken with a 14\arcsec\ slit, which is large compared to the typical seeing-limited point-spread function (PSF) of the star ($\sim1\arcsec$).  Because the slit width is comparable to the seeing-limited PSF, the HIRES PSF varies over time with changes in seeing, weather, and aberrations of the wavefront. The CPS Doppler routine involves forward-modeling the iodine-imprinted spectrum of a star as a combination of a high-resolution iodine library spectrum and an iodine-free, velocity-shifted, PSF-deconvolved template spectrum of the target star. The deconvolved template spectrum is obtained by observing rapidly-rotating B stars with the iodine cell in the light path immediately before and after an iodine-free observation of the star, which effectively samples the PSF at the time of the template.  The individual spectra with iodine are then convolved with a PSF of HIRES that is determined for each observation. These new RVs are presented in Table \ref{tab:rvs}.

\section{Orbit Analysis and Planet Properties}

\begin{figure*}
    \centering
    \includegraphics[width=0.65\textwidth]{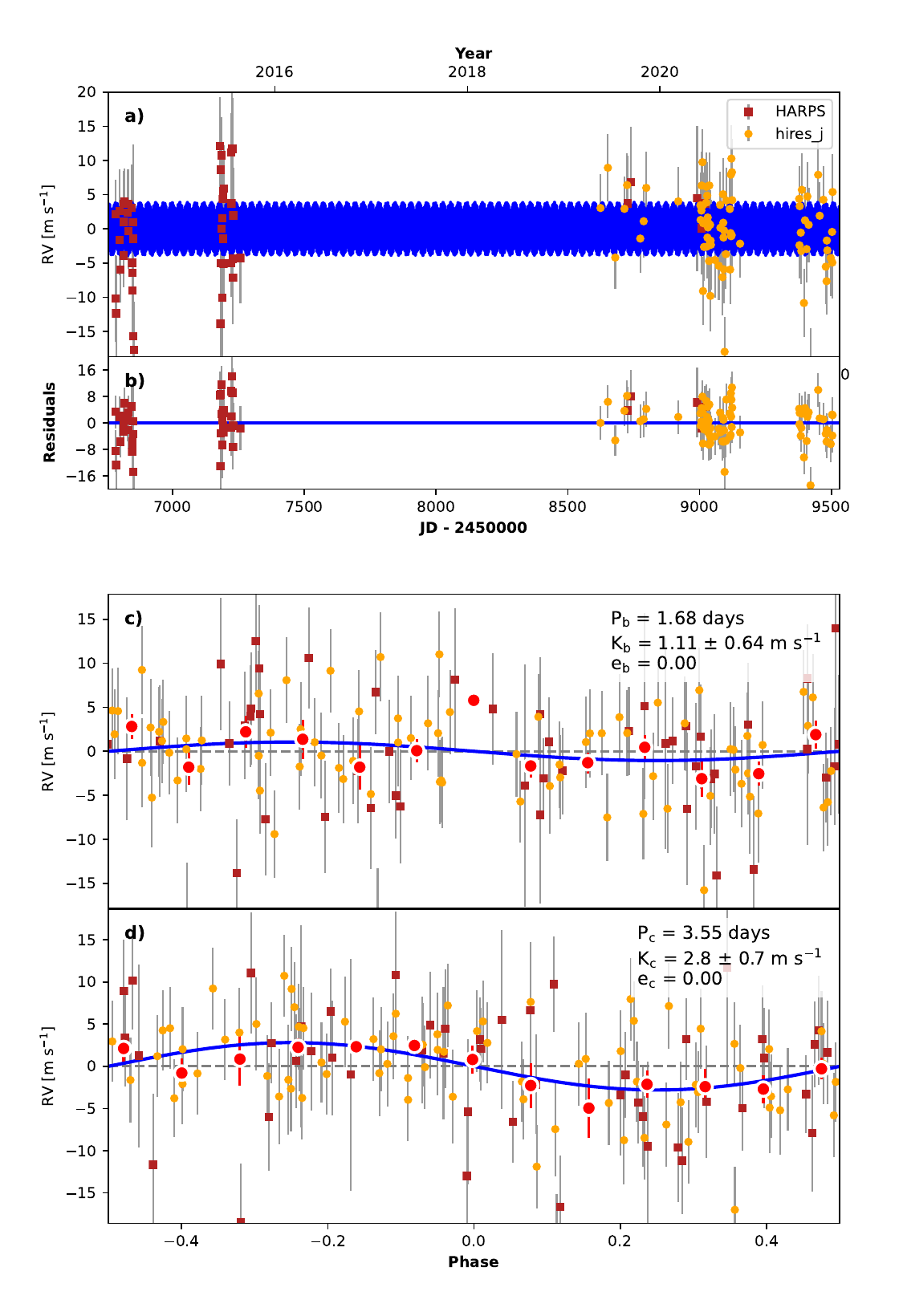}\hfill
    \caption{\textbf{Top:} RVs for Kepler-323, from HIRES (yellow) over 80 nights between May 2019 and October 2021 combined with 48 RVs from HARPS-N (dark red) \citep{Bonomo2023}. The blue line represents the best fit Keplerian orbits. The residuals are shown in panel b). \textbf{Bottom:} Phase folded RVs and the corresponding best-fit Keplerian orbit for each individual planet. The red circles are the RVs binned in orbital phase.}
    \label{fig:K323_RV}
\end{figure*}

\begin{figure*}
    \centering
    \includegraphics[width=0.65\textwidth]{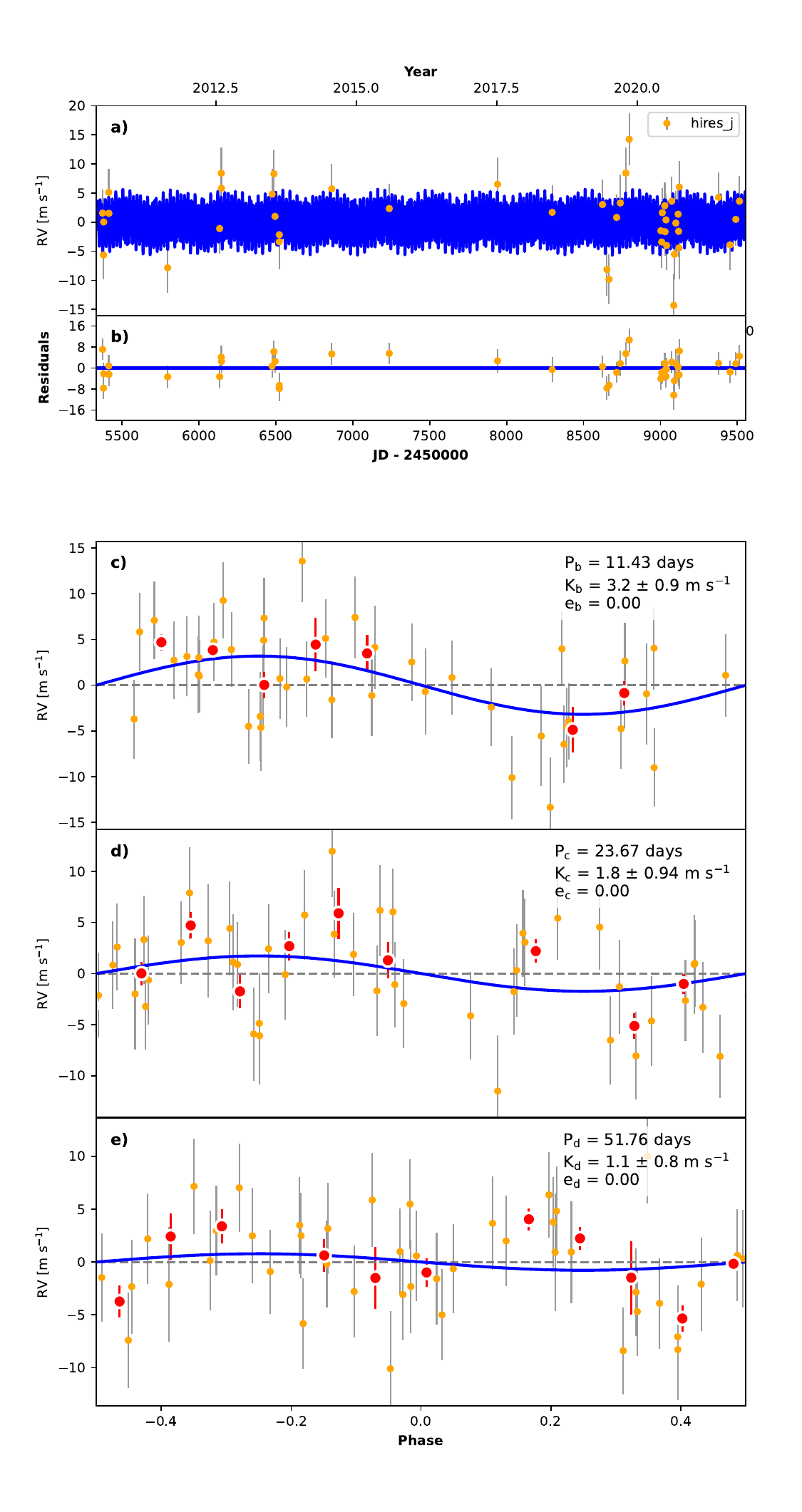}\hfill
    \caption{Same as Figure \ref{fig:K323_RV}, but for Kepler-104} 
    \label{fig:K104_RV}
\end{figure*}

We used the open source python package {\fontfamily{qcr}\selectfont RadVel} \citep{Fulton2018} to model our RVs with a Keplerian orbit. We set the periods and time of conjunctions for all five planets to the values reported in \citet{Morton2016}, which are presented in Table \ref{tab:orbits}. 
Because the RV semi-amplitudes were small compared to the per-measurement error, making it difficult to accurately determine the eccentricities, we fixed the eccentricity of all planets at zero. Our choice to fix the eccentricities at zero is further motivated by the observation that most small planets in multi-planet systems have small eccentricities (median $e=0.02$, \citealt{Yee2021}). Thus, our only free parameters were the RV semi-amplitudes, RV zero-point ($\gamma$), and RV jitter ($\sigma_i$) for each instrument.  We adopted priors ensuring that the RV semi-amplitude and the jitter were positive. We performed a maximum likelihood fit using {\fontfamily{qcr}\selectfont RadVel}. To determine the uncertainties in our parameters, we used {\fontfamily{qcr}\selectfont RadVel}'s Markov Chain Monte Carlo (MCMC) method with 500 walkers and 100,000 steps. The maximum Gelman-Rubin value for the convergence of our MCMC is $<$ 1.001 for both Kepler-323 and Kepler-104. The resulting confidence intervals for each parameter are displayed in Table \ref{tab:orbits}. The RVs and the best-fit models are shown in Figures \ref{fig:K323_RV} and \ref{fig:K104_RV}.

\begin{figure*}
    \centering
    \includegraphics[width=\textwidth]{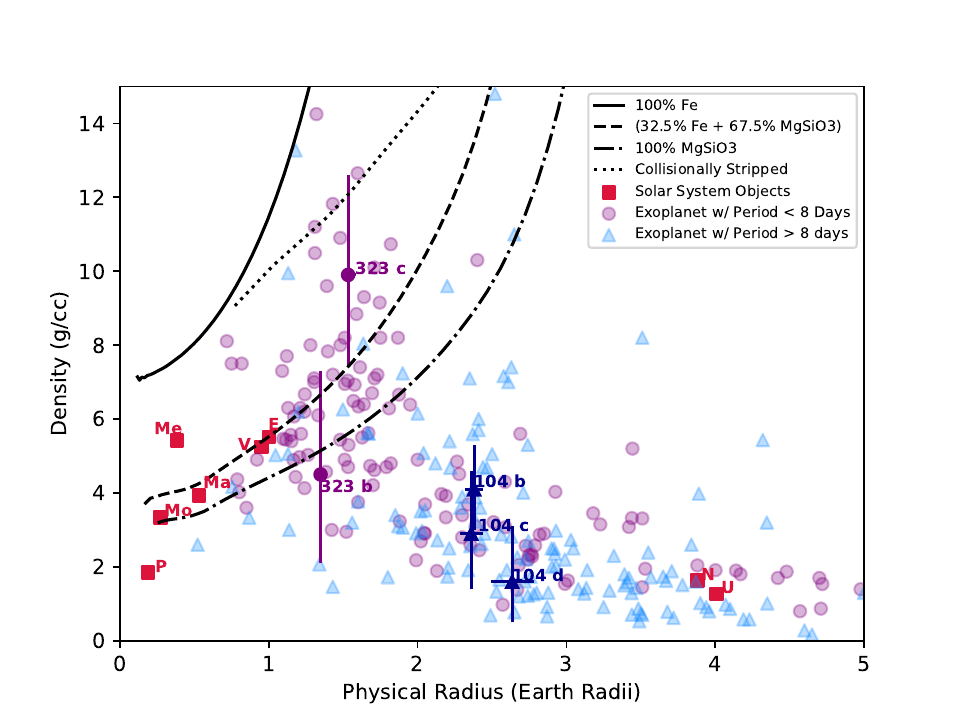}\hfill
    \caption{Planet densities vs physical radii for Kepler-104 (dark blue), Kepler-323 (dark purple), Solar System objects (red squares), and planets selected from the NASA Exoplanet Archive according to Appendix \ref{sec:appendix} (translucent). The Solar System objects include Mercury, Venus, Earth, Mars, Uranus, Neptune, Pluto and the Moon. Non-Solar System objects are colored by their orbital period (see legend). The composition curves consistent with 100\% Fe, 100\% MgSiO$_{3}$, and 67.5\% MgSiO$_{3}$ + 32.5\% Fe are taken from \cite{Zeng2019}, while the maximal collisionally stripped composition curve is taken from \citet{Marcus2010}.} 

    \label{fig:Density_vs_Radius}
\end{figure*}

\begin{figure*}
    \centering
    \includegraphics[width=\textwidth]{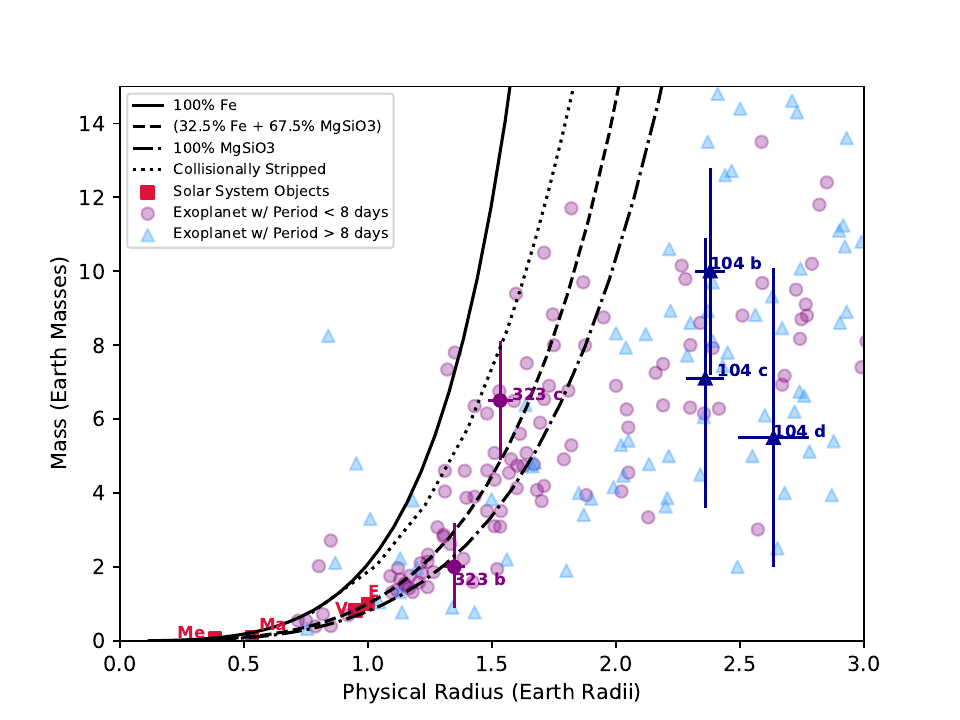}\hfill
    \caption{Same as Figure \ref{fig:Density_vs_Radius}, but with the y-axis corresponding to mass instead of density.} 
    \label{fig:Mass_vs_Radius}
\end{figure*}

Our Keplerian orbital fit yields masses of $2.0^{+1.2}_{-1.1}$ M$_\Earth$ and 6.5$\pm1.6$ M$_\Earth$ for Kepler-323 b and c and masses of 10.0$\pm2.8$ M$_\Earth$, $7.1^{+3.8}_{-3.5}$ M$_\Earth$, and $5.5^{+4.6}_{-3.5}$ M$_\Earth$ for Kepler-104 b, c, and d (Figures \ref{fig:Density_vs_Radius} and \ref{fig:Mass_vs_Radius}). We compare these to a sample of known exoplanets, selected by the process described in Appendix \ref{sec:appendix}, and objects from the Solar System. In Figures \ref{fig:Density_vs_Radius} and \ref{fig:Mass_vs_Radius}, we focused solely on planets with calculated densities; we removed any planets with a mass derived from a mass radius relation. Figures \ref{fig:Density_vs_Radius} and \ref{fig:Mass_vs_Radius} include lines of constant composition and collisionally stripped planets \citep{Zeng2019,Marcus2010}. 

Kepler-323 c's density of $9.9^{+2.7}_{-2.5}$ g/cc is consistent with a rocky composition. The 1$\sigma$ credible interval spans an Earth-like composition or a more iron-rich planet. Kepler-323 b, however, has a density of $4.5^{+2.8}_{-2.4}$ g/cc, which is lower than what we would expect for a planet of pure silicates.  One possible interpretation of a low density is that the planet could have a gaseous envelope of high mean molecular weight (such as has been suggested for TOI-561 b in \citealt{Brinkman2023}). However, the 1$\sigma$ confidence interval spans densities of purely silicate and Earth-like compositions.

The architecture of the Kepler-323 system provides an interesting example to test atmospheric mass loss processes. Planets b and c have similar radii, but their masses appear to differ by a factor of 3, with the outer planet being more massive. The short orbital period of Kepler-323 c, combined with a density consistent with a rocky composition, indicates that it would be unlikely to retain a gaseous envelope in the presence of intense stellar X-ray and ultraviolet radiation from the star\citep{OwenSchlichting23}. In addition, we would expect Kepler-323 c to lose its atmosphere through core-powered mass loss due to the short orbital period and high equilibrium temperature \citep{Ginzburg2018}. If we assume that Kepler-323 c, which is rocky, lost its envelope, then Kepler-323 b, which is closer to its star and (apparently) lower mass, should also have lost its envelope. Follow-up observations of Kepler-323 b are needed to better constrain its mass and composition.

The three Kepler-104 planets, with densities of $4.1^{+1.2}_{-1.1}$, $2.9^{+1.7}_{-1.5}$ g/cc, and $1.6^{+1.5}_{-1.1}$ g/cc respectively, are all consistent with bulk compositions that include a gaseous envelope. Rocky compositions are ruled out for all three planets with $>3\sigma$ confidence, which is not surprising given that their radii are all $>2\,R_\oplus$ \citep{Fulton2017,Rogers2015,Weiss2014}.

To explore the influence of orbital period on planet masses and radii, in Figures \ref{fig:Density_vs_Radius} and \ref{fig:Mass_vs_Radius} we split our sample of planets with $R< 5 R_\Earth$ into ``short period'' ($P < 8$ days, purple circles) and ``long period'' ($P > 8$ days, blue triangles) groups \footnote{This is an integer near the median orbital period of our sample: 7.97 days}. For planets that are ``short period,'' 59 of 134, including Kepler-323, have a best-fit density that is consistent with a rocky composition. For the ``long-period'' group, 119 planets of 130, including Kepler-104, have best-fit densities and radii that indicate a gaseous envelope. 

There are at least two plausible explanations for this pattern: one astrophysical, and one related to selection bias.  The prevalence of gas envelopes at longer orbital periods vs. bare rocky planets at short periods is consistent with predictions from atmospheric mass-loss models \citep{OwenSchlichting23,Ginzburg2018}. For photoevaporation, planets at longer periods are less irradiated by stellar X-ray and UV radiation, meaning they are less likely to lose their envelopes. For core-powered mass loss, hotter equilibrium temperatures increases the odds of losing an atmosphere from the cooling of the planet. A mass-radius-period relationship was developed in \citet{MillsMazeh2017}. In addition to these astrophysical phenomenon, the relationship between planet mass, radius, and orbital period is sculpted, at least partially, by selection bias.  For instance, the masses of small planets ($R < 1.5 R_\oplus$) are typically only measured when the planet is sufficiently close to the star to produce a detectable RV amplitude, and so the masses of small planets at long orbital periods have rarely been measured.  To further clarify the relative impact of selection biases and atmospheric mass-loss on sculpting this relationship, more observations of small planets at long periods are required. Transit timing variations provide a viable method for measuring the masses of some rocky planets at $P > 8$ days and will yield additional insights on the relationship between planet mass, radius, and orbital period, especially in multi-planet systems e.g. \citep{Agol21}. 

\section{Planetary System Properties}
\subsection{Mass Partitioning}
In addition to contextualizing the properties of the individual planets, we can characterize these planetary systems with metrics of their architectures.  Mass partitioning and gap complexity describe the mass diversity and spacing distribution of a system, respectively. These metrics have been used to evaluate both observed and synthetic planetary system architectures \citep{GF2020}. 

The mass partitioning of a system is given by:

\begin{equation}
\mathcal{Q} \equiv \biggl(\frac{N}{N-1}\biggl)\cdot\biggl(\sum_{i=1}^{N} \biggl( m_{i}^{*} - \frac{1}{N}\biggl)^{2}\biggl) 
\end{equation}
where
\begin{equation}
m^{*}_{i} = m_{i}/\sum^{N}_{i}m_{i}
\end{equation}

Here, $N$ is the number of planets in a system and $m_{i}$ is the mass of the $i^{\mathrm{th}}$ planet (order does not matter). The normalizing coefficient, $\biggl(\frac{N}{N-1}\biggl)$, restricts $\mathcal{Q}$ to be from 0 to 1, where a value of 0 indicates equal masses throughout the system and a value that approaches 1 represents a system with a dominant giant planet and $N-1$ tiny planets. Note that mass partitioning is agnostic to the ordering of the planets. 

\begin{figure*}
    \centering
    \includegraphics[width=\textwidth]{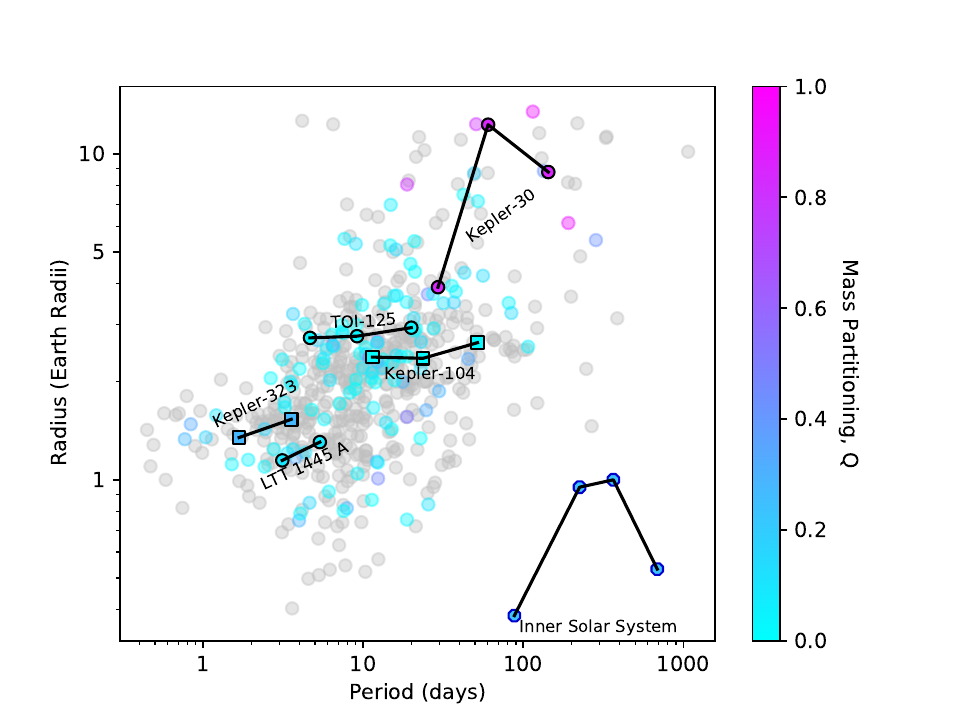}\hfill
    \caption{Individual planet radii against their orbital periods and colored by their mass partitioning \citep{GF2020} for 649 planets (229 planetary systems). For planets colored gray (459 planets in 121 systems), we do not provide an estimate for the mass partitioning because the system has a planet missing from the steps outlined in Appendix \ref{sec:appendix}. In addition to our measurements for Kepler-323 and Kepler-104 (squares), LTT 1445 A, TOI-125, Kepler-30 and the Inner Solar System are highlighted for comparison.}
    \label{fig:MP}
\end{figure*}

Figure \ref{fig:MP} shows the planet radius (R$_\Earth$) and orbital period  (days) for Kepler-323, Kepler-104, and other small exoplanets selected from the \citet{NASAExoArch}; Appendix \ref{sec:appendix}).  Each planet is colored by the mass partitioning for its system, based on the masses of all the known planets in that system. 
In order for a system to appear in Figure \ref{fig:MP} in color, it must have all planets survive the selection process outlined in Appendix \ref{sec:appendix} and it must have a mass measurement for every known planet in the system. Otherwise, the system is colored gray; an example is Kepler-11 as Kepler-11 c's mass error is too large for our selection criteria and Kepler-11 g does not have a measured mass.

\begin{figure*}
    \centering
    \includegraphics[width=\textwidth]{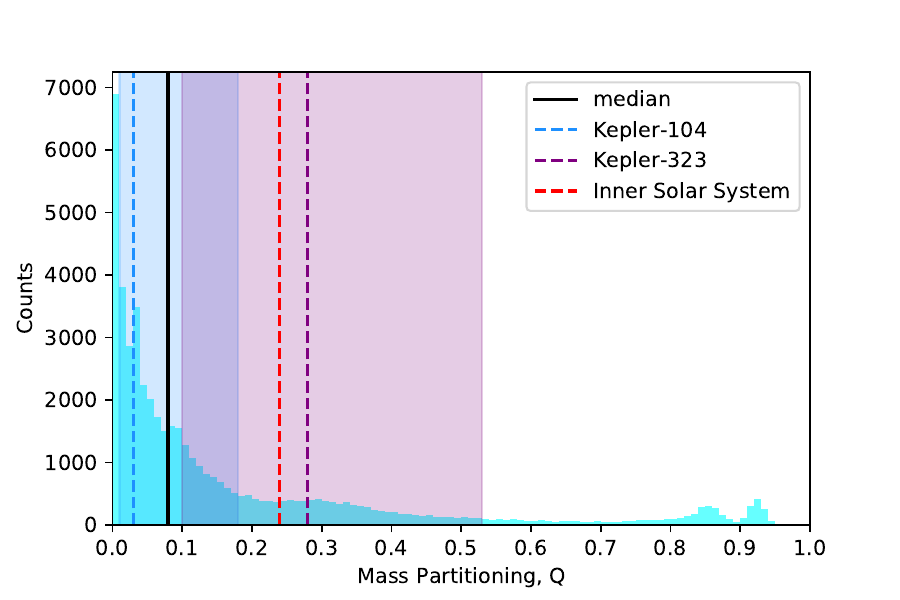}\hfill
    \caption{The posterior distribution of mass partitioning values, based on 1000 draws from the posteriors of the planet masses, for the systems described in Section 4.1. The solid black line represents the median mass partitioning of the sample. The dashed lines represent the best fit mass partitioning value for Kepler-104 (blue), Kepler-323 (purple), and the Inner Solar System (red). The shaded regions represent the 1$\sigma$ errors.}
    \label{fig:allpartition}
\end{figure*}

We calculated the distribution of the possible mass partitioning values for each system in our sample based on the mass measurement values and errors. For each system we selected planet masses from a normal distribution defined by the masses and errors reported on the NASA Exoplanet Archive.  We used rejection sampling to discard any negative mass draws.  We iterated until each planet had 1000 draws from our reconstructed mass posterior and calculated the mass partitioning for each iteration.  This resulted in 49,000 mass partitioning values across 49 systems, the distribution of which is shown in Figure \ref{fig:allpartition}. The median mass partitioning in this sample is $\mathcal{Q} = 0.08$.

 Kepler-104, with a mass partitioning of $\mathcal{Q} = 0.03$, is consistent with many of the systems in our sample (median $\mathcal{Q} = 0.06$ for three-planet systems). Kepler-323, on the other hand, has the twelfth highest mass partitioning in our 52 system sample ($\mathcal{Q} = 0.28$), which is slightly larger than the Inner Solar System ($\mathcal{Q} = 0.24$) and is higher than typical (median $\mathcal{Q} = 0.10$ for two-planet systems). \footnote{We define the Inner Solar System as Mercury, Venus, Earth, and Mars, the region consistent with the $\lesssim 1$ a.u. orbits typical of transiting exoplanets}. In order to probe the relationship between the mass diversity and architecture of a multi-planet system, we highlight LTT 1445 A (\citealt{Winters2022, Pass2023}) and TOI-125 (\citealt{Nielsen2020}) in Figure \ref{fig:MP} as systems with similar architectures to Kepler-323 and Kepler-104 respectively. TOI-125, which has three planets of similar radii to Kepler-104 but with shorter orbital periods, has a mass partitioning of $\mathcal{Q} = 0.04$, similar to both Kepler-104 and the median mass partitioning for three-planet systems in our sample. LTT 1445 A, a two-planet system with smaller radii and longer orbits than Kepler-323, has a mass partitioning  of $\mathcal{Q} = 0.11$, which is typical for the two-planet systems and lower than the mass partition of Kepler-323.  While Kepler-323 has a higher mass partitioning than typical, it is not a strong example of a high mass partitioning system; Kepler-30 \citep{Sanchis-Ojeda2012}, a system with two Neptune-mass planets and one super-Jupiter, has a mass partitioning of $\mathcal{Q} = 0.85$.  A more detailed treatment of the distribution of mass partitioning in various planetary architectures would be enlightening, but is beyond the scope of this paper.

\begin{figure*}
    \centering
    \includegraphics[width=\textwidth]{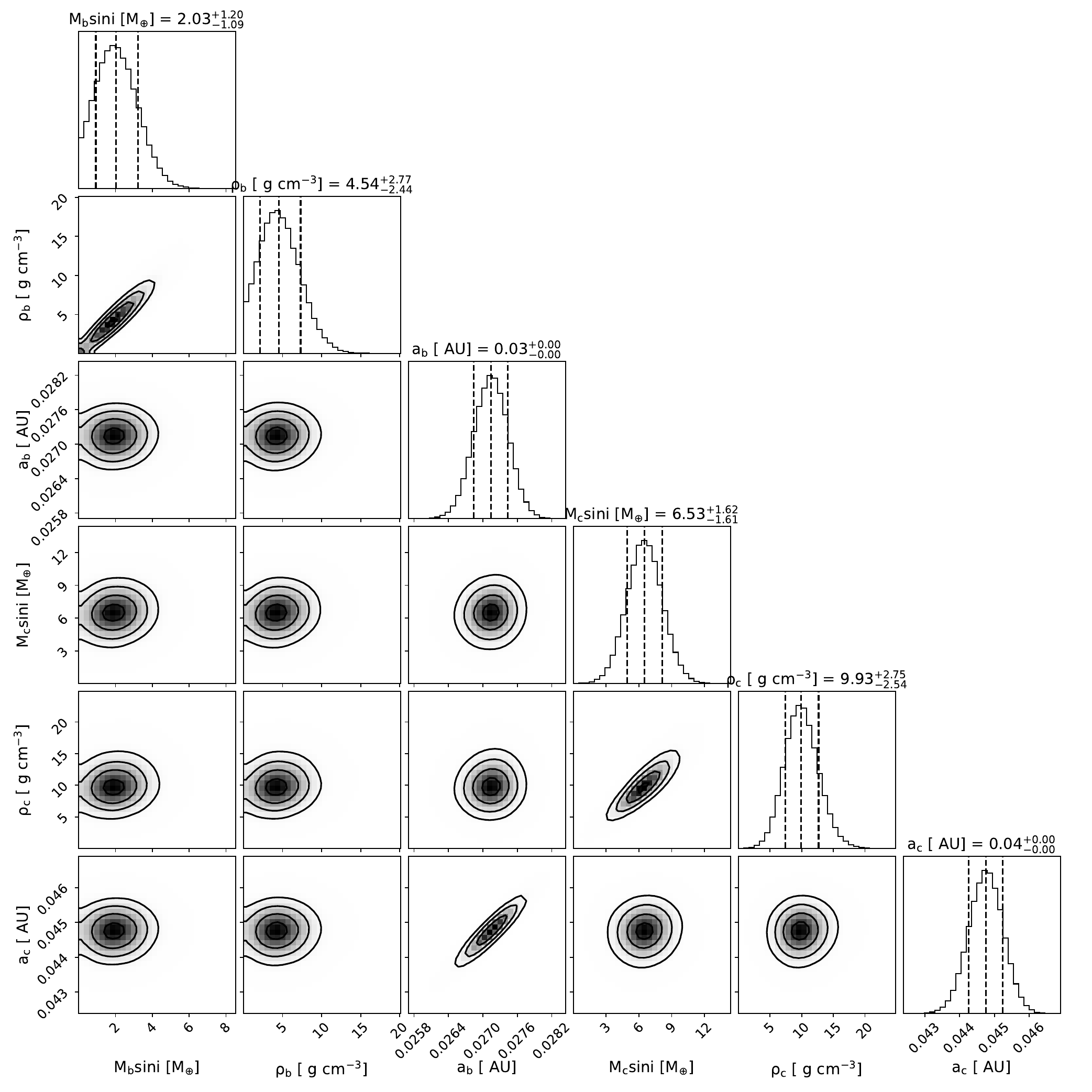}\hfill
    \caption{Posterior distributions for all derived parameters from the Keplerian best-fit orbit of Kepler-323. There is no strong covariance between the masses of planets b and c}
    \label{fig:Corner323}
\end{figure*}

\begin{figure*}
    \centering
    \includegraphics[width=\textwidth]{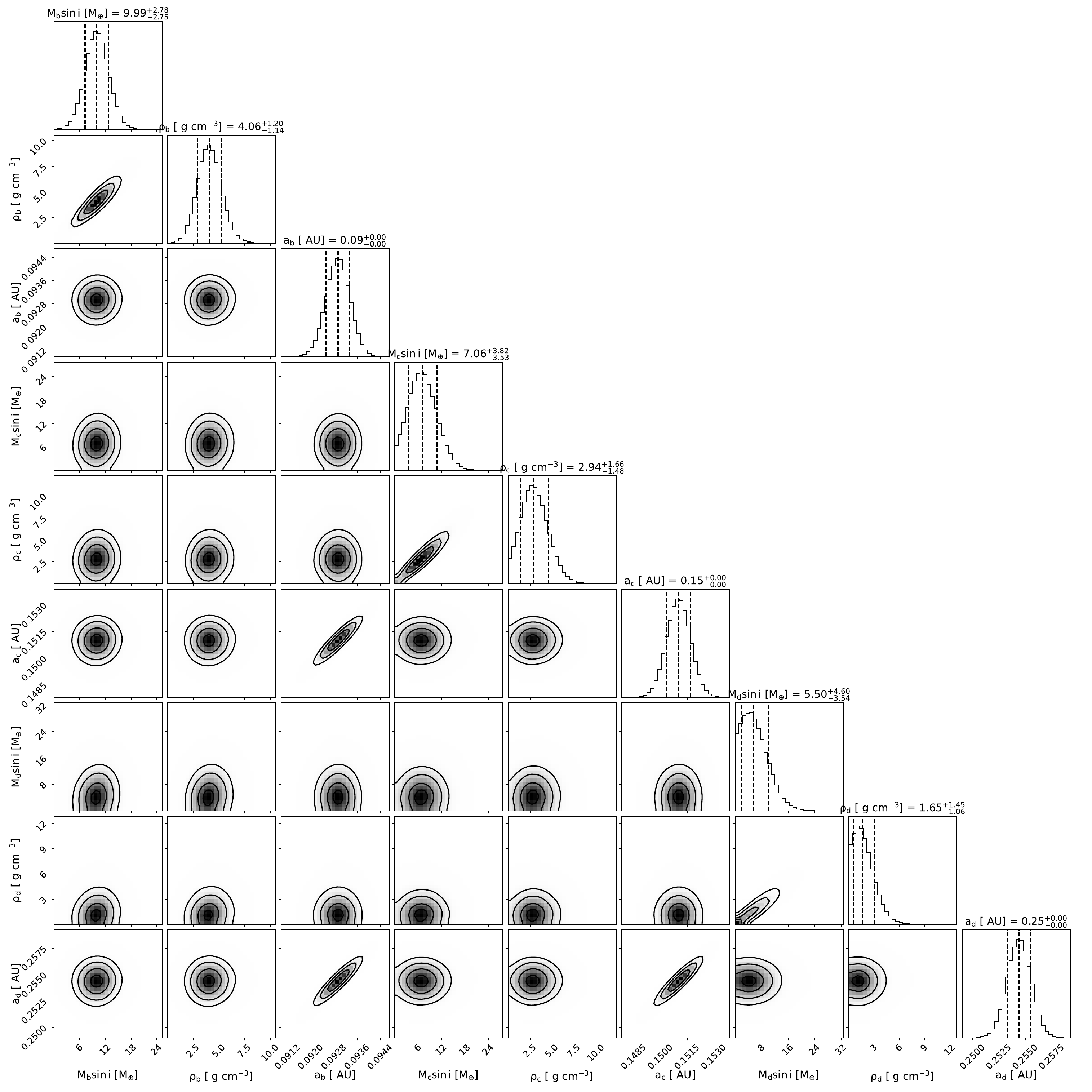}\hfill
    \caption{Same as Figure \ref{fig:Corner323}, but for Kepler-104.}
    \label{fig:Corner104}
\end{figure*}

\begin{figure*}
    \centering
    \includegraphics[width=\textwidth]{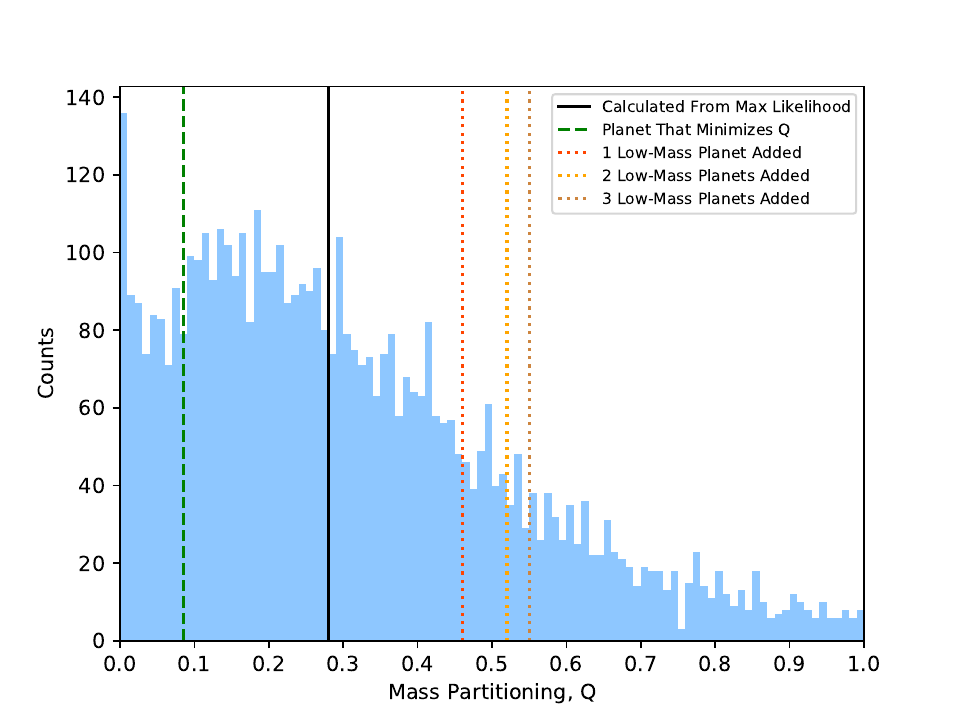}\hfill
    \caption{The posterior distribution of mass partitioning values for the Kepler-323 system, based on 5000 draws from the posteriors of the planet masses. The solid black line represents the calculated mass partitioning from the maximum likelihood mass probabilities. The three dotted lines represent the mass partitioning value in hypothetical scenarios where one (orange), two (gold), or three (brown) planets of zero mass are also present in the system. The green dashed line represents another hypothetical scenario, where there is a third planet with a mass of $5.5\,M_\Earth$, which minimizes the mass partitioning value.  The peak in the posterior near zero is real.}
    \label{fig:K323_hist}
\end{figure*}

\begin{figure*}
    \centering
    \includegraphics[width=\textwidth]{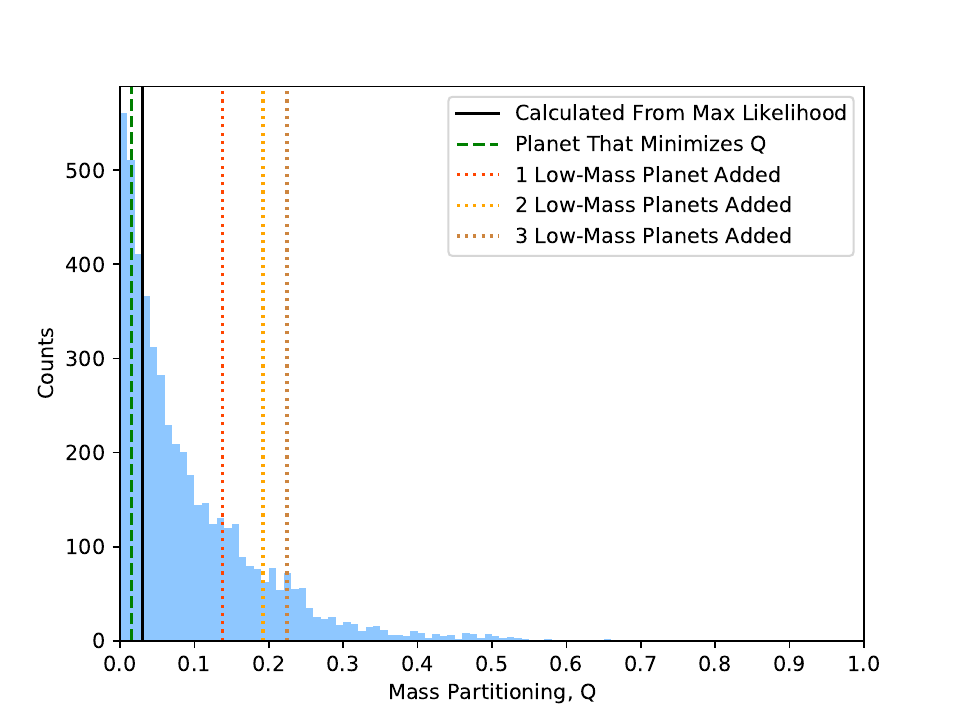}\hfill
    \caption{Same as Figure \ref{fig:K323_hist}, but for Kepler-104}
    \label{fig:K104_hist}
\end{figure*}

We investigated how the errors on the planet masses propagate to the mass partitioning of the system to determine the accuracy of the calculated mass partitioning and our confidence in the characterization of the mass diversity. We selected 5000 possible mass combinations for Kepler-323 and Kepler-104 from the masses defined by the mass and error for each planet\footnote{Formally, it would be better to draw from the joint posterior distribution, but we see no covariance between the planet masses (Figures \ref{fig:Corner323},\ref{fig:Corner104})}  and used rejection sampling to ensure all masses were positive. For each combination of planet masses, we calculated the mass partitioning. Figure \ref{fig:K323_hist} and Figure \ref{fig:K104_hist} show histograms of these calculated possible mass partitioning values for Kepler-323 and Kepler-104 respectively. Each plot has a solid black line to indicate the calculated median value of the credible interval. 

\begin{figure*}
    \centering
    \includegraphics[width=\textwidth]{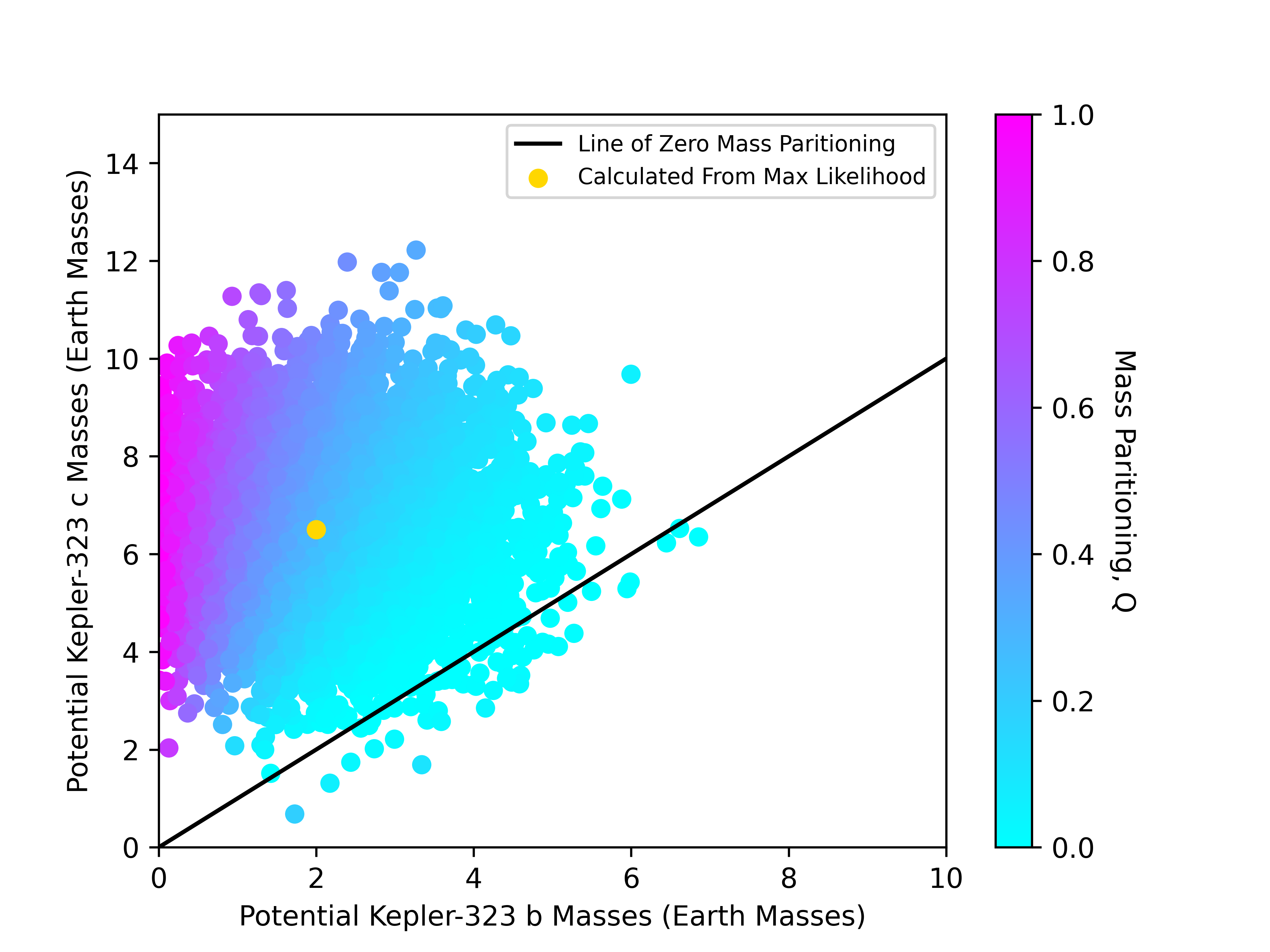}\hfill
    \caption{Potential masses of Kepler-323 b vs potential masses of Kepler-323 c for 5000 random draws from the posteriors of the individual planets, colored by the mass partitioning \citep{GF2020}. The yellow dot shows the calculated value from the maximum likelihood values and the black line represents a mass partitioning of exactly zero.}
    \label{fig:K323 Error}
\end{figure*}

The Kepler-323 mass partitioning posterior spans the full range of zero to one as the error bars on the planet masses are large. There is a peak in the posterior at a mass partitioning near zero due to more cases where Kepler-323 b and Kepler-323 c have very similar masses than very different masses, as seen in Figure \ref{fig:K323 Error}. 

It is possible that there could be additional, undetected planets in a system. This produces another source of uncertainty in our assessment of the mass partitioning because a planet with significantly different mass from the detected planets could strongly affect the mass partitioning. Figures \ref{fig:K323_hist} and \ref{fig:K104_hist} have lines that represent the mass partitioning for the system if one (orange), two (gold), or three (brown) zero-mass planets are undetected in the system. To explore the opposite extreme (an undetected planet of similar mass to the known planets), we added an extra planet and varied its mass to minimize the mass partitioning. The mass of this extra planet is 5.5 M$\Earth$ for Kepler-323 and 8.0 M$\Earth$ for Kepler-104. The mass partitioning for all of these possibilities are included in Table \ref{tab:mass_partitioning}. The presence of a Jupiter in either Kepler-323 or Kepler-104 would be intriguing as it would dramatically impact the mass partitioning of the system. Based on a decade of radial velocity measurements of Kepler-323, we have no evidence of the presence of a Jupiter out to 10 AU and we find a 3$\sigma$ upper limit of 1 $M_{J}$. The presence of a Jupiter in Kepler-104 was ruled out in \citet{KGPS}.

For Kepler-104, 92\% of the mass partitioning posterior is below the mass partitioning of the inner Solar System, 98\% is below the mass partitioning of the gaseous Solar System ($\mathcal{Q} = 0.41$), and 99\% is below the mass partitioning of the full Solar System ($\mathcal{Q} = 0.49$). Even if Kepler-104 has three undetected planets of zero mass, its mass partitioning is below that of the inner Solar System. We find that the Kepler-104 mass partitioning, compared to Kepler-323, is well-constrained and that there is low mass diversity in this system.

\subsection{Gap Complexity\label{sec:gc}}
 
Another metric we calculated to analyze the architectures of multi-planet systems is the gap complexity \citep{GF2020}. Gap complexity characterizes the irregularity of the orbital periods of planets in a single system:

\begin{equation}
\mathcal{C} = -K\biggl(\sum_{i=1}^{n} p_{i}^{*}\log p_{i}^{*}\biggl)\cdot\biggl(\sum_{i=1}^{n}\biggl(p_{i}^{*}-\frac{1}{n}\biggl)^{2}\biggl) 
\end{equation}
where
$K$ is a normalization constant,
$n$ is the number of gaps between planets,
$P^{'}$ and $P$ are the outer and inner periods of pairs of adjacent planets, and
$p_{i}^{*}$ are defined as the ``pseudo-probabilities'' and are normalized to one: 
\begin{equation}
p^*_i \equiv \frac{\log(P^{'}/P)}{\log(P_{\max}/P_{\min})}
\end{equation}
where $P_{\max}$ and $P_{\min}$ are respectively defined as the maximum period and the minimum period of any planet in the system. Like mass partitioning, gap complexity ranges from zero (evenly spaced planets) to one (maximum $p_{i}^{*} \approx \frac{2}{3}$).  

The gap complexity of Kepler-104, along with a sample of known multi-planet systems from the NASA Exoplanet Archive (see Appendix \ref{sec:appendix}), is shown in Figure \ref{fig:GC}. In order to calculate the gap complexity, there must be at least three planets in the system. Any system with only two known planets, such as Kepler-323 and LTT 1445 A, is colored gray. In addition, any system that has a planet that was removed from the data set by the process described in Appendix \ref{sec:appendix} is also colored gray. Unlike Figure \ref{fig:MP}, we do not exclude planets with a mass derived from a mass-radius relation. Kepler-104 ($\mathcal{C} = 0.004$) and TOI-125 ($\mathcal{C} = 0.01$) have low gap complexities, indicating that they are almost perfectly evenly spaced. In contrast, Kepler-167 \citep{Chachan2022} has high gap complexity ($\mathcal{C} = 0.80 $). Kepler-104's low gap complexity, which is the 6th lowest among the systems examined here (66 total systems), is consistent with the other exoplanet systems, whereas systems with architectures like that of Kepler-167 are rarely detected. The inner Solar System has a gap complexity of $\mathcal{C} = 0.13$ and has been noted for its regular spacing \citep{Nieto1970}.

\section{Conclusion}
We conducted an analysis of 79 newly gathered RVs, combined with 48 RVs from \citet{Bonomo2023}, of Kepler-323 and 44 RVs from \citet{KGPS} of Kepler-104 in order to quantify the masses and densities of their planets. Using these newly calculated masses, we computed the mass diversity and period spacing of these systems and compared those results to other multi-planet transiting systems. We also compared the individual planet masses, radii, and period to planets from other multi-planet transiting systems.

Based on the orbital periods and planet radii of the Kepler-323 planets, we find that these planets are, mostly, as we would expect when compared to the general exoplanet population. In addition, when looking at the relationship between orbital period and physical radius, 79\% of planets in our sample with radii less than 1.8 R$_\Earth$ have an orbital period less than 8 days. This includes the Kepler-323 planets.

The calculated mass partitioning of Kepler-323 is $0.28^{+0.25}_{-0.18}$, which is 0.04 larger than the Inner Solar System and is the twelfth largest in our sample. Due to the errors on the individual planet masses, the mass partitioning posterior for Kepler-323 contains values from 0 to 1, which covers the entire range of possible values for mass partitioning. In addition, the presence of undetected planets could also shift the calculated mass partitioning value by a factor of two. Since we cannot calculate a gap complexity for systems with only two planets, we cannot comment on the spacing of the Kepler-323 system. 

The Kepler-104 planets, based on their radii and orbital periods, are similar to the other exoplanets with an orbital period of longer than 8 days have radii larger than 1.8 R$_\Earth$ shown in Figure \ref{fig:Density_vs_Radius} and Figure \ref{fig:Mass_vs_Radius}.

Kepler-104 has a low mass partitioning of $\mathcal{Q}=0.03^{+0.15}_{-0.02}$. If there were three undetected (zero mass) planets in the Kepler-104 system, the mass partitioning for Kepler-104 would remain lower than that of the inner Solar System. This indicates that Kepler-104 has very little mass diversity between the planets. The gap complexity of Kepler-104 is 0.004, which is the 6th lowest of the 66 systems in our sample with three or more planets and can have a calculated gap complexity. This further reinforces Kepler-104 as a "peas-in-a-pod" system.

In this paper, we considered how the following questions applied to the peas-in-a-pod systems of Kepler-323 and Kepler-104: 1) What are the masses and bulk compositions of the individual planets? (2) Are the masses and radii of the planets typical for their orbital periods? (3) Within a given system, what is the mass diversity and spacing distribution of transiting planets? (4) How could the presence of undetected, low-mass planets affect the statistics of the intra-system size and mass uniformity? While Kepler-323 and Kepler-104 both display radius uniformity, they do not share the same level of mass uniformity as Kepler-104 appears to have much more uniform masses. Applying these questions to a larger sample of high multiplicity systems that exhibit radius uniformity or diversity will provide valuable constraints for planet formation and will serve as a basis for future work.

\begin{acknowledgements}
We acknowledge support in the form of observational resources at W. M. Keck Observatory from NASA and the University of Hawai`i.  L.M.W. acknowledges support from the NASA-Keck Key Strategic Mission Support program (grant no. 80NSSC19K1475) and the NASA Exoplanet Research Program (grant no. 80NSSC23K0269).  

This work was supported by a NASA Keck PI Data Award, administered by the NASA Exoplanet Science Institute. Data presented herein were obtained at the W. M. Keck Observatory from telescope time allocated to the National Aeronautics and Space Administration through the agency's scientific partnership with the California Institute of Technology and the University of California. The Observatory was made possible by the generous financial support of the W. M. Keck Foundation.  This dataset made use of the NASA Exoplanet Science Institute at IPAC, which is operated by the California Institute of Technology under contract with the National Aeronautics and Space Administration.

We acknowledge contributions from B.J. Fulton for maintaining the Jump database in which data for this project were curated.  We acknowledge L. A. Rogers for contributing to the proposals that led to the collection of RVs presented here and for useful discussions and comments on the manuscript.

We would like to thank Greg Gilbert for investigating the presence of TTVs in these systems.

The authors wish to recognize and acknowledge the very significant cultural role and reverence that the summit of Maunakea has always had within the indigenous Hawaiian community. We are most fortunate to have the opportunity to conduct observations from this mountain.
\end{acknowledgements}

\facility{Keck:I (HIRES)}

\software{RADVEL (Fulton et al. 2018)}

\clearpage
\appendix
\section{NASA Exoplanet Archive Sample Selection\label{sec:appendix}}
In order to place the Kepler-323 and Kepler-104 planets into the context of the general exoplanet population, we used a curated sample of planets from the NASA Exoplanet Archive Confirmed Planets Table as comparison:

\begin{enumerate}
    \item We started with the largest possible sample by downloading all of the confirmed planets from the NASA Exoplanet Archive, as of November 8th, 2022.  We included all the rows from the database, which sometimes resulted in multiple entries for a single planet.
    \item We removed each row that did not include an entry for the orbital period, planet physical radius, or planet physical radius error.
    \item We removed each row that had a non-transiting flag (indicating a non-transiting planet).
    \item We removed each row that had a controversial flag.
    \item We removed each row that had a mass value but no mass error, as these rows corresponded to mass upper limits.
    \item We removed rows where the planet masses were determined from \citet{Hadden2014} because the masses from that study were sometimes biased \citep{Hadden2017}.
    \item Following these cuts, we sorted by the NASA Exoplanet Archive's default parameter. If a planet still had an entry in our sample but the NASA Exoplanet Archive-determined ``default'' row was already dropped, we sorted by the date of update. We then removed any duplicate row, keeping either the default parameter set or the most recent data for each planet.
    \item For rows that did not have an entry for planet mass, we used the mass radius relation from \citet{Weiss2014} to estimate a mass. This was done to preserve the system's architectures when calculating gap complexity.
    \item For rows that did not have an entry for planet density, we computed the planet density. 
    \item We calculated the percent error for the radius, mass, and density of each planet. We then removed planets with a radius error $>$10\%, a mass error $>$85\%, or a density error $>$104\%. These figures were chosen based on the calculated values and errors of Kepler-323 and Kepler-104. We wanted to ensure that we selected planets with as precise measurements as possible, while being sure that our planets would not be excluded from this population. 
\end{enumerate}
\clearpage

\section{RV Table}
\centering

\startlongtable
\begin{deluxetable}{ccc}
\tablecaption{HIRES RVs of Kepler-323\label{tab:rvs}}
\tablehead{\colhead{Time} & \colhead{RV} & \colhead{RV Error} \\ 
\colhead{(BJD-2450000 )} & \colhead{(m/s)} & \colhead{(m/s)} } 
\startdata
8624.079912 &   2.69 & 2.80  \\
8652.076300 &   8.57 & 2.79  \\
8679.936895 &  -4.56 & 2.09  \\
8714.811564 &   2.57 & 2.25  \\
8724.815753 &   6.04 & 2.28  \\
8774.870351 &  -1.78 & 2.92  \\
8787.831759 &   0.75 & 3.43  \\
8797.796716 &   5.63 & 3.35  \\
8919.108473 &   3.66 & 2.98  \\
9003.034178 &   3.07 & 2.46  \\
9003.943462 &   0.94 & 2.73  \\
9006.929700 &   2.32 & 2.82  \\
9007.965294 &   5.97 & 2.68  \\
9010.977440 &   9.39 & 2.50  \\
9011.964465 &   3.06 & 2.52  \\
9012.940900 &  -9.46 & 2.80  \\
9013.914590 &   0.47 & 2.54  \\
9016.896354 &  -3.01 & 2.74  \\
9024.905287 &   3.46 & 2.51  \\
9025.987166 &   5.80 & 2.76  \\
9027.872397 &  -2.96 & 2.28  \\
9028.839619 &   4.50 & 2.67  \\
9029.901252 &  -1.64 & 2.32  \\
9030.934017 &   1.31 & 2.70  \\
9031.878713 &  -0.15 & 2.62  \\
9034.919478 &   0.42 & 2.42  \\
9035.940978 &  -5.07 & 2.73  \\
9036.819639 &   5.98 & 2.58  \\
9038.913723 &  -2.15 & 2.48  \\
9039.908818 &   0.23 & 2.95  \\
9041.058788 & -10.17 & 3.19  \\
9041.854713 &  -2.08 & 2.69  \\
9042.981926 &  -2.69 & 5.08  \\
9057.995943 &  -4.91 & 3.75  \\
9072.969537 &  -5.76 & 2.80  \\
9077.945189 &   3.14 & 2.54  \\
9078.961976 &  -0.20 & 3.03  \\
9086.924455 &  -7.42 & 3.86  \\
9088.943940 &   4.70 & 3.37  \\
9089.937189 &   0.53 & 2.52  \\
9090.875231 &  -1.63 & 2.43  \\
9091.831883 &  -6.28 & 2.82  \\
9092.896238 &  -0.61 & 2.41  \\
9094.904686 & -18.34 & 2.99  \\
9099.855638 &  -4.06 & 2.97  \\
9101.846391 &   3.78 & 2.41  \\
9114.782555 &   2.55 & 2.80  \\
9115.894902 &  -6.32 & 2.83  \\
9117.798061 &  -0.99 & 2.49  \\
9118.789695 &   7.57 & 2.82  \\
9119.808424 &  -1.43 & 3.14  \\
9120.799605 &   9.92 & 2.61  \\
9121.829744 &  -1.28 & 2.72  \\
9122.828842 &   7.92 & 2.64  \\
9123.756542 &   3.83 & 2.42  \\
9153.731898 &  -2.53 & 2.78  \\
9377.958508 &   4.03 & 2.50  \\
9379.032190 &  -2.76 & 2.51  \\
9379.985243 &   2.96 & 2.44  \\
9384.046876 &  -0.91 & 2.60  \\
9386.055153 &  -3.60 & 2.55  \\
9387.978216 &   5.34 & 3.45  \\
9396.017499 & -11.21 & 2.76  \\
9399.979627 &   0.88 & 2.59  \\
9406.054052 &  -3.32 & 2.70  \\
9409.077191 &   4.38 & 2.95  \\
9414.024875 &   0.23 & 3.56  \\
9420.991303 & -20.42 & 3.72  \\
9449.967276 &   7.55 & 2.95  \\
9455.879048 &   1.56 & 2.73  \\
9469.855090 &   3.90 & 2.75  \\
9478.887201 &  -5.90 & 3.55  \\
9481.833186 &  -8.02 & 2.67  \\
9482.889332 &  -2.07 & 2.90  \\
9484.799051 &  -3.52 & 2.43  \\
9497.827107 &  -4.62 & 3.82  \\
9502.837147 &  -0.81 & 2.72  \\
9503.815084 &  -5.29 & 3.20  \\
9504.824258 &   5.03 & 3.68  \\
\enddata

\tablecomments{These values come from the HIRES data alone and are presented before any gamma offset or jitter have been applied. Times are in BJD-TDB.}
\end{deluxetable}

\begin{figure*}
    \centering
    \includegraphics[width=\textwidth]{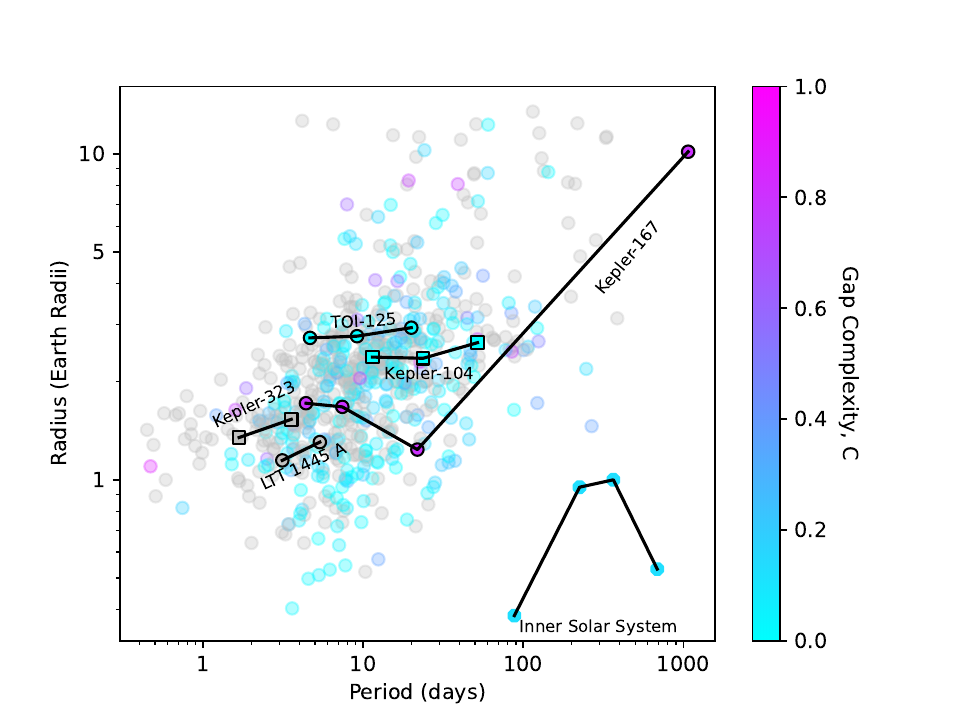}\hfill
    \caption{Individual planet radii against their orbital periods and colored by their gap complexity \citep{GF2020} for 649 planets (229 planetary systems). For planets colored gray (353 planets in 161 systems), we cannot provide an estimate for the gap complexity because either the system only has two planets, of which gap complexity is undefined, or the system has a planet missing due to the steps outlined in Appendix \ref{sec:appendix}. In addition to our measurements for Kepler-323 and Kepler-104 (squares), LTT 1445 A, TOI-125, Kepler-167 and the Inner Solar System are highlighted for comparison.}
    \label{fig:GC}
\end{figure*}

\begin{table*}[]
\caption{\label{tab:orbits} Planet Orbital Properties}

\begin{center}
\begin{tabular}{l|l|l|l|}
\cline{2-4}
                                  & $P$ (days) & $T_c$ (BJD) & $K (m/s)$ \\ \hline
\multicolumn{1}{|l|}{Kepler-323 b} &$1.67832545 \pm 0.000002$ & $2454969.052935 \pm 0.0009$ & $1.37^{+0.73}_{-0.69}$ \\ \hline
\multicolumn{1}{|l|}{Kepler-323 c} &$3.553829671 \pm 0.000005$ & $2454967.64145 \pm 0.001$ & $2.78\pm 0.74$ \\ \hline
\multicolumn{1}{|l|}{Kepler-104 b} &$11.427545631 \pm 0.000009$ & $2454970.613679 \pm 0.0006$ & $3.24^{+0.91}_{-0.89}$\\ \hline
\multicolumn{1}{|l|}{Kepler-104 c} &$23.66836427 \pm 0.00003$ & $2454965.71366 \pm 0.001$ & $1.79^{+0.98}_{-0.91}$ \\ \hline
\multicolumn{1}{|l|}{Kepler-104 d} & $51.75529576 \pm 0.00008$ & $2455104.08787 \pm 0.001$ & $1.07^{+0.92}_{-0.7}$ \\ \hline
\end{tabular}

\tablecomments{The eccentricities and arguments of periastron were taken as zero for each planet. The $\gamma$ and jitter for Kepler-104 are $-0.88^{+0.73}_{-0.74}$ m/s and $3.73^{+0.71}_{0.63}$ m/s respectively. For Kepler-323, the $\gamma$ and jitter from HIRES are $-0.36\pm0.58$ m/s and $4.26^{+0.53}_{-0.48}$ m/s and the $\gamma$ and jitter from HARPS-N are $-2.8^{0.96}_{-0.97}$ m/s and $2.1^{+1.6}_{-1.4}$ m/s.}

\end{center}
\end{table*}

\begin{table*}[]
\caption{Planet Physical Properties \label{tab:physical}}
\begin{center}
\begin{tabular}{l|l|l|l|}
\cline{2-4}
                                  & Msini (M$\Earth$) & Radius (R$\Earth$) & Density (g/cc) \\ \hline
\multicolumn{1}{|l|}{Kepler-323 b} & $2.0^{+1.2}_{-1.1}$ & $1.348\pm 0.042$ & $4.5^{+2.8}_{-2.4}$ \\ \hline
\multicolumn{1}{|l|}{Kepler-323 c} & $6.5\pm 1.6$ & $1.533\pm 0.050$ & $9.9^{+2.7}_{-2.5}$ \\ \hline
\multicolumn{1}{|l|}{Kepler-104 b} & $10.0\pm 2.8$ & $2.380\pm 0.060$ & $4.1^{+1.2}_{-1.1}$ \\ \hline
\multicolumn{1}{|l|}{Kepler-104 c} & $7.1^{+3.8}_{-3.5}$ & $2.360\pm 0.077$ & $2.9^{+1.7}_{-1.5}$ \\ \hline
\multicolumn{1}{|l|}{Kepler-104 d} & $5.5^{+4.6}_{-3.5}$ & $2.635\pm 0.144$ & $1.6^{+1.5}_{-1.1}$ \\ \hline
\end{tabular}
\end{center}
\end{table*}

\begin{table*}[]
\caption{Values of Mass Partitioning\label{tab:mass_partitioning}}
\begin{center}
\begin{tabular}{l|l|l|l|l|l|}
\cline{2-6}
                                  & Best-Fit & Minimized w/Extra Planet & 1 Massless Planet & 2 Massless Planets & 3 Massless Planets \\ \hline
\multicolumn{1}{|l|}{Kepler-323} & $0.28^{+0.25}_{-0.18}$ & 0.09 & 0.46 & 0.52 & 0.55\\ \hline
\multicolumn{1}{|l|}{Kepler-104} & $0.03^{+0.15}_{-0.02}$ & 0.02 & 0.14 & 0.19 & 0.22\\ \hline
\end{tabular}
\end{center}
\end{table*}
\clearpage

\bibliography{main}{} 

\begin{thebibliography}{}
\expandafter\ifx\csname natexlab\endcsname\relax\def\natexlab#1{#1}\fi
\providecommand{\url}[1]{\href{#1}{#1}}
\providecommand{\dodoi}[1]{doi:~\href{http://doi.org/#1}{\nolinkurl{#1}}}
\providecommand{\doeprint}[1]{\href{http://ascl.net/#1}{\nolinkurl{http://ascl.net/#1}}}
\providecommand{\doarXiv}[1]{\href{https://arxiv.org/abs/#1}{\nolinkurl{https://arxiv.org/abs/#1}}}

\bibitem[{{Agol} {et~al.}(2021){Agol}, {Dorn}, {Grimm}, {Turbet}, {Ducrot}, {Delrez}, {Gillon}, {Demory}, {Burdanov}, {Barkaoui}, {Benkhaldoun}, {Bolmont}, {Burgasser}, {Carey}, {de Wit}, {Fabrycky}, {Foreman-Mackey}, {Haldemann}, {Hernandez}, {Ingalls}, {Jehin}, {Langford}, {Leconte}, {Lederer}, {Luger}, {Malhotra}, {Meadows}, {Morris}, {Pozuelos}, {Queloz}, {Raymond}, {Selsis}, {Sestovic}, {Triaud}, \& {Van Grootel}}]{Agol21}
{Agol}, E., {Dorn}, C., {Grimm}, S.~L., {et~al.} 2021, Planetary Science Journal, 2, 1, \dodoi{10.3847/PSJ/abd022}

\bibitem[{{Bonomo} {et~al.}(2023){Bonomo}, {Dumusque}, {Massa}, {Mortier}, {Bongiolatti}, {Malavolta}, {Sozzetti}, {Buchhave}, {Damasso}, {Haywood}, {Morbidelli}, {Latham}, {Molinari}, {Pepe}, {Poretti}, {Udry}, {Affer}, {Boschin}, {Charbonneau}, {Cosentino}, {Cretignier}, {Ghedina}, {Lega}, {L{\'o}pez-Morales}, {Margini}, {Mart{\'\i}nez Fiorenzano}, {Mayor}, {Micela}, {Pedani}, {Pinamonti}, {Rice}, {Sasselov}, {Tronsgaard}, \& {Vanderburg}}]{Bonomo2023}
{Bonomo}, A.~S., {Dumusque}, X., {Massa}, A., {et~al.} 2023, arXiv e-prints, arXiv:2304.05773, \dodoi{10.48550/arXiv.2304.05773}

\bibitem[{{Brinkman} {et~al.}(2023){Brinkman}, {Weiss}, {Dai}, {Huber}, {Kite}, {Valencia}, {Bean}, {Beard}, {Behmard}, {Blunt}, {Brady}, {Fulton}, {Giacalone}, {Howard}, {Isaacson}, {Kasper}, {Lubin}, {MacDougall}, {Akana Murphy}, {Plotnykov}, {Polanski}, {Rice}, {Seifahrt}, {Stef{\'a}nsson}, \& {St{\"u}rmer}}]{Brinkman2023}
{Brinkman}, C.~L., {Weiss}, L.~M., {Dai}, F., {et~al.} 2023, \aj, 165, 88, \dodoi{10.3847/1538-3881/acad83}

\bibitem[{{Butler} {et~al.}(1996){Butler}, {Marcy}, {Williams}, {McCarthy}, {Dosanjh}, \& {Vogt}}]{Butler1996}
{Butler}, R.~P., {Marcy}, G.~W., {Williams}, E., {et~al.} 1996, \pasp, 108, 500, \dodoi{10.1086/133755}

\bibitem[{{Chachan} {et~al.}(2022){Chachan}, {Dalba}, {Knutson}, {Fulton}, {Thorngren}, {Beichman}, {Ciardi}, {Howard}, \& {Van Zandt}}]{Chachan2022}
{Chachan}, Y., {Dalba}, P.~A., {Knutson}, H.~A., {et~al.} 2022, \apj, 926, 62, \dodoi{10.3847/1538-4357/ac3ed6}

\bibitem[{{Fulton} \& {Petigura}(2018)}]{CKS7}
{Fulton}, B.~J., \& {Petigura}, E.~A. 2018, \aj, 156, 264, \dodoi{10.3847/1538-3881/aae828}

\bibitem[{{Fulton} {et~al.}(2018){Fulton}, {Petigura}, {Blunt}, \& {Sinukoff}}]{Fulton2018}
{Fulton}, B.~J., {Petigura}, E.~A., {Blunt}, S., \& {Sinukoff}, E. 2018, \pasp, 130, 044504, \dodoi{10.1088/1538-3873/aaaaa8}

\bibitem[{{Fulton} {et~al.}(2017){Fulton}, {Petigura}, {Howard}, {Isaacson}, {Marcy}, {Cargile}, {Hebb}, {Weiss}, {Johnson}, {Morton}, {Sinukoff}, {Crossfield}, \& {Hirsch}}]{Fulton2017}
{Fulton}, B.~J., {Petigura}, E.~A., {Howard}, A.~W., {et~al.} 2017, \aj, 154, 109, \dodoi{10.3847/1538-3881/aa80eb}

\bibitem[{{Furlan} \& {Howell}(2017)}]{Furlan2017}
{Furlan}, E., \& {Howell}, S.~B. 2017, \aj, 154, 66, \dodoi{10.3847/1538-3881/aa7b70}

\bibitem[{{Gaia Collaboration}(2020)}]{GAIADR3}
{Gaia Collaboration}. 2020, VizieR Online Data Catalog, I/350

\bibitem[{{Gilbert} \& {Fabrycky}(2020)}]{GF2020}
{Gilbert}, G.~J., \& {Fabrycky}, D.~C. 2020, \aj, 159, 281, \dodoi{10.3847/1538-3881/ab8e3c}

\bibitem[{{Ginzburg} {et~al.}(2018){Ginzburg}, {Schlichting}, \& {Sari}}]{Ginzburg2018}
{Ginzburg}, S., {Schlichting}, H.~E., \& {Sari}, R. 2018, \mnras, 476, 759, \dodoi{10.1093/mnras/sty290}

\bibitem[{{Hadden} \& {Lithwick}(2014)}]{Hadden2014}
{Hadden}, S., \& {Lithwick}, Y. 2014, \apj, 787, 80, \dodoi{10.1088/0004-637X/787/1/80}

\bibitem[{{Hadden} \& {Lithwick}(2017)}]{Hadden2017}
---. 2017, \aj, 154, 5, \dodoi{10.3847/1538-3881/aa71ef}

\bibitem[{{Lillo-Box} {et~al.}(2014){Lillo-Box}, {Barrado}, \& {Bouy}}]{Lillo-Box2014}
{Lillo-Box}, J., {Barrado}, D., \& {Bouy}, H. 2014, \aap, 566, A103, \dodoi{10.1051/0004-6361/201423497}

\bibitem[{{Marcus} {et~al.}(2010){Marcus}, {Sasselov}, {Hernquist}, \& {Stewart}}]{Marcus2010}
{Marcus}, R.~A., {Sasselov}, D., {Hernquist}, L., \& {Stewart}, S.~T. 2010, \apjl, 712, L73, \dodoi{10.1088/2041-8205/712/1/L73}

\bibitem[{{Marcy} {et~al.}(2014){Marcy}, {Isaacson}, {Howard}, {Rowe}, {Jenkins}, {Bryson}, {Latham}, {Howell}, {Gautier}, {Batalha}, {Rogers}, {Ciardi}, {Fischer}, {Gilliland}, {Kjeldsen}, {Christensen-Dalsgaard}, {Huber}, {Chaplin}, {Basu}, {Buchhave}, {Quinn}, {Borucki}, {Koch}, {Hunter}, {Caldwell}, {Van Cleve}, {Kolbl}, {Weiss}, {Petigura}, {Seager}, {Morton}, {Johnson}, {Ballard}, {Burke}, {Cochran}, {Endl}, {MacQueen}, {Everett}, {Lissauer}, {Ford}, {Torres}, {Fressin}, {Brown}, {Steffen}, {Charbonneau}, {Basri}, {Sasselov}, {Winn}, {Sanchis-Ojeda}, {Christiansen}, {Adams}, {Henze}, {Dupree}, {Fabrycky}, {Fortney}, {Tarter}, {Holman}, {Tenenbaum}, {Shporer}, {Lucas}, {Welsh}, {Orosz}, {Bedding}, {Campante}, {Davies}, {Elsworth}, {Handberg}, {Hekker}, {Karoff}, {Kawaler}, {Lund}, {Lundkvist}, {Metcalfe}, {Miglio}, {Silva Aguirre}, {Stello}, {White}, {Boss}, {Devore}, {Gould}, {Prsa}, {Agol}, {Barclay}, {Coughlin}, {Brugamyer}, {Mullally}, {Quintana}, {Still}, {Thompson}, {Morrison}, {Twicken},
  {D{\'e}sert}, {Carter}, {Crepp}, {H{\'e}brard}, {Santerne}, {Moutou}, {Sobeck}, {Hudgins}, {Haas}, {Robertson}, {Lillo-Box}, \& {Barrado}}]{Marcy2014}
{Marcy}, G.~W., {Isaacson}, H., {Howard}, A.~W., {et~al.} 2014, \apjs, 210, 20, \dodoi{10.1088/0067-0049/210/2/20}

\bibitem[{{Mills} \& {Mazeh}(2017)}]{MillsMazeh2017}
{Mills}, S.~M., \& {Mazeh}, T. 2017, \apjl, 839, L8, \dodoi{10.3847/2041-8213/aa67eb}

\bibitem[{{Morton} {et~al.}(2016){Morton}, {Bryson}, {Coughlin}, {Rowe}, {Ravichandran}, {Petigura}, {Haas}, \& {Batalha}}]{Morton2016}
{Morton}, T.~D., {Bryson}, S.~T., {Coughlin}, J.~L., {et~al.} 2016, \apj, 822, 86, \dodoi{10.3847/0004-637X/822/2/86}

\bibitem[{{Mugrauer}(2019)}]{Mugrauer2019}
{Mugrauer}, M. 2019, \mnras, 490, 5088, \dodoi{10.1093/mnras/stz2673}

\bibitem[{{Mulders} {et~al.}(2018){Mulders}, {Pascucci}, {Apai}, \& {Ciesla}}]{Mulders2018}
{Mulders}, G.~D., {Pascucci}, I., {Apai}, D., \& {Ciesla}, F.~J. 2018, \aj, 156, 24, \dodoi{10.3847/1538-3881/aac5ea}

\bibitem[{{NASA Exoplanet Archive}(2019)}]{NASAExoArch}
{NASA Exoplanet Archive}. 2019, Confirmed Planets Table,  IPAC, \dodoi{10.26133/NEA1}

\bibitem[{{Nielsen} {et~al.}(2020){Nielsen}, {Gandolfi}, {Armstrong}, {Jenkins}, {Fridlund}, {Santos}, {Dai}, {Adibekyan}, {Luque}, {Steffen}, {Esposito}, {Meru}, {Sabotta}, {Bolmont}, {Kossakowski}, {Otegi}, {Murgas}, {Stalport}, {Rodler}, {D{\'\i}az}, {Kurtovic}, {Ricker}, {Vanderspek}, {Latham}, {Seager}, {Winn}, {Jenkins}, {Allart}, {Almenara}, {Barrado}, {Barros}, {Bayliss}, {Berdi{\~n}as}, {Boisse}, {Bouchy}, {Boyd}, {Brown}, {Bryant}, {Burke}, {Cochran}, {Cooke}, {Demangeon}, {D{\'\i}az}, {Dittman}, {Dorn}, {Dumusque}, {Garc{\'\i}a}, {Gonz{\'a}lez-Cuesta}, {Grziwa}, {Georgieva}, {Guerrero}, {Hatzes}, {Helled}, {Henze}, {Hojjatpanah}, {Korth}, {Lam}, {Lillo-Box}, {Lopez}, {Livingston}, {Mathur}, {Mousis}, {Narita}, {Osborn}, {Palle}, {Rojas}, {Persson}, {Quinn}, {Rauer}, {Redfield}, {Santerne}, {dos Santos}, {Seidel}, {Sousa}, {Ting}, {Turbet}, {Udry}, {Vanderburg}, {Van Eylen}, {Vines}, {Wheatley}, \& {Wilson}}]{Nielsen2020}
{Nielsen}, L.~D., {Gandolfi}, D., {Armstrong}, D.~J., {et~al.} 2020, \mnras, 492, 5399, \dodoi{10.1093/mnras/staa197}

\bibitem[{{Nieto}(1970)}]{Nieto1970}
{Nieto}, M.~M. 1970, \aap, 8, 105

\bibitem[{{Owen} \& {Schlichting}(2023)}]{OwenSchlichting23}
{Owen}, J.~E., \& {Schlichting}, H.~E. 2023, arXiv e-prints, arXiv:2308.00020, \dodoi{10.48550/arXiv.2308.00020}

\bibitem[{{Pass} {et~al.}(2023){Pass}, {Winters}, {Charbonneau}, {Balkanski}, {Lewis}, {Lally}, {Bean}, {Cloutier}, \& {Eastman}}]{Pass2023}
{Pass}, E.~K., {Winters}, J.~G., {Charbonneau}, D., {et~al.} 2023, arXiv e-prints, arXiv:2307.02970.
\newblock \doarXiv{2307.02970}

\bibitem[{{Rogers}(2015)}]{Rogers2015}
{Rogers}, L.~A. 2015, \apj, 801, 41, \dodoi{10.1088/0004-637X/801/1/41}

\bibitem[{{Sanchis-Ojeda} {et~al.}(2012){Sanchis-Ojeda}, {Fabrycky}, {Winn}, {Barclay}, {Clarke}, {Ford}, {Fortney}, {Geary}, {Holman}, {Howard}, {Jenkins}, {Koch}, {Lissauer}, {Marcy}, {Mullally}, {Ragozzine}, {Seader}, {Still}, \& {Thompson}}]{Sanchis-Ojeda2012}
{Sanchis-Ojeda}, R., {Fabrycky}, D.~C., {Winn}, J.~N., {et~al.} 2012, \nat, 487, 449, \dodoi{10.1038/nature11301}

\bibitem[{{Vogt} {et~al.}(1994){Vogt}, {Allen}, {Bigelow}, {Bresee}, {Brown}, {Cantrall}, {Conrad}, {Couture}, {Delaney}, {Epps}, {Hilyard}, {Hilyard}, {Horn}, {Jern}, {Kanto}, {Keane}, {Kibrick}, {Lewis}, {Osborne}, {Pardeilhan}, {Pfister}, {Ricketts}, {Robinson}, {Stover}, {Tucker}, {Ward}, \& {Wei}}]{Vogt1994}
{Vogt}, S.~S., {Allen}, S.~L., {Bigelow}, B.~C., {et~al.} 1994, in Society of Photo-Optical Instrumentation Engineers (SPIE) Conference Series, Vol. 2198, Instrumentation in Astronomy VIII, ed. D.~L. {Crawford} \& E.~R. {Craine}, 362, \dodoi{10.1117/12.176725}

\bibitem[{{Weiss} \& {Marcy}(2014)}]{Weiss2014}
{Weiss}, L.~M., \& {Marcy}, G.~W. 2014, \apjl, 783, L6, \dodoi{10.1088/2041-8205/783/1/L6}

\bibitem[{{Weiss} {et~al.}(2018){Weiss}, {Marcy}, {Petigura}, {Fulton}, {Howard}, {Winn}, {Isaacson}, {Morton}, {Hirsch}, {Sinukoff}, {Cumming}, {Hebb}, \& {Cargile}}]{Weiss2018}
{Weiss}, L.~M., {Marcy}, G.~W., {Petigura}, E.~A., {et~al.} 2018, \aj, 155, 48, \dodoi{10.3847/1538-3881/aa9ff6}

\bibitem[{{Weiss} {et~al.}(2024){Weiss}, {Isaacson}, {Howard}, {Fulton}, {Petigura}, {Fabrycky}, {Jontof-Hutter}, {Steffen}, {Schlichting}, {Wright}, {Beard}, {Brinkman}, {Chontos}, {Giacalone}, {Hill}, {Kosiarek}, {MacDougall}, {Mo{\v{c}}nik}, {Polanski}, {Turtelboom}, {Tyler}, \& {Van Zandt}}]{KGPS}
{Weiss}, L.~M., {Isaacson}, H., {Howard}, A.~W., {et~al.} 2024, \apjs, 270, 8, \dodoi{10.3847/1538-4365/ad0cab}

\bibitem[{{Winters} {et~al.}(2022){Winters}, {Cloutier}, {Medina}, {Irwin}, {Charbonneau}, {Astudillo-Defru}, {Bonfils}, {Howard}, {Isaacson}, {Bean}, {Seifahrt}, {Teske}, {Eastman}, {Twicken}, {Collins}, {Jensen}, {Quinn}, {Payne}, {Kristiansen}, {Spencer}, {Vanderburg}, {Zechmeister}, {Weiss}, {Wang}, {Wang}, {Udry}, {Terentev}, {St{\"u}rmer}, {Stef{\'a}nsson}, {Shporer}, {Shectman}, {Sefako}, {Schwengeler}, {Schwarz}, {Scarsdale}, {Rubenzahl}, {Roy}, {Rosenthal}, {Robertson}, {Petigura}, {Pepe}, {Omohundro}, {Murphy}, {Murgas}, {Mo{\v{c}}nik}, {Montet}, {Mennickent}, {Mayo}, {Massey}, {Lubin}, {Lovis}, {Lewin}, {Kasper}, {Kane}, {Jenkins}, {Huber}, {Horne}, {Hill}, {Gorrini}, {Giacalone}, {Fulton}, {Forveille}, {Figueira}, {Fetherolf}, {Dressing}, {D{\'\i}az}, {Delfosse}, {Dalba}, {Dai}, {Cort{\'e}s}, {Crossfield}, {Crane}, {Conti}, {Collins}, {Chontos}, {Butler}, {Brown}, {Brady}, {Behmard}, {Beard}, {Batalha}, \& {Almenara}}]{Winters2022}
{Winters}, J.~G., {Cloutier}, R., {Medina}, A.~A., {et~al.} 2022, \aj, 163, 168, \dodoi{10.3847/1538-3881/ac50a9}

\bibitem[{{Yee} {et~al.}(2021){Yee}, {Tamayo}, {Hadden}, \& {Winn}}]{Yee2021}
{Yee}, S.~W., {Tamayo}, D., {Hadden}, S., \& {Winn}, J.~N. 2021, \aj, 162, 55, \dodoi{10.3847/1538-3881/ac00a9}

\bibitem[{{Zeng} {et~al.}(2019){Zeng}, {Jacobsen}, {Sasselov}, {Petaev}, {Vanderburg}, {Lopez-Morales}, {Perez-Mercader}, {Mattsson}, {Li}, {Heising}, {Bonomo}, {Damasso}, {Berger}, {Cao}, {Levi}, \& {Wordsworth}}]{Zeng2019}
{Zeng}, L., {Jacobsen}, S.~B., {Sasselov}, D.~D., {et~al.} 2019, Proceedings of the National Academy of Science, 116, 9723, \dodoi{10.1073/pnas.1812905116}

\bibitem[{{Zhu} {et~al.}(2018){Zhu}, {Petrovich}, {Wu}, {Dong}, \& {Xie}}]{Zhu2018}
{Zhu}, W., {Petrovich}, C., {Wu}, Y., {Dong}, S., \& {Xie}, J. 2018, \apj, 860, 101, \dodoi{10.3847/1538-4357/aac6d5}

\end{thebibliography}
\bibliographystyle{aasjournal}

\end{document}